 \documentclass[nohyper,12pt,letterpaper]{JHEP}
 \usepackage{epsfig}
 \newfont{\frak}{eufm10 scaled 1200}
 
 \newfont{\Bbb}{msbm10 scaled 1200} %instead of eusb10
 \newcommand{\mathbb}[1]{\mbox{\Bbb #1}}
 \DeclareSymbolFont{AMSa}{U}{msa}{m}{n}
 \DeclareSymbolFont{AMSb}{U}{msb}{m}{n}
 \let\Box\relax
 \DeclareMathSymbol{\Box}{\mathord}{AMSa}{"03}

 \def\Ref#1{(\ref{#1})}
 \def \eqn#1#2{\begin{equation}#2\label{#1}\end{equation}}

 \title{M-theory observables for cosmological space-times}
 \author{T.\,Banks\thanks{On leave from Rutgers U.}\\
 Department of Physics \\
 University of California, Santa Cruz, CA 95064\\
 E-mail: \email{banks@scipp.ucsc.edu}}
 \author{W.\,Fischler\\
 Department of Physics\\
 University of Texas, Austin, TX 78712\\
 E-mail: \email{fischler@physics.utexas.edu}}
 \abstract{We discuss the construction of the analog of an S-matrix for
 space-times that begin with a Big-Bang and asymptote to an FRW
 universe with nonnegative cosmological constant. When the
 cosmological constant is positive there are many such S-matrices,
 related mathematically by gauge transformations and physically by an
 analog of the principle of black hole complementarity. In the limit
 of vanishing $\Lambda$ these become (approximate) Poincare transforms
 of each other. Considerations of the initial state require a quantum
 treatment of space-time, and some preliminary steps towards
 constructing such a theory are proposed. In this context we propose
 a model for the earliest semiclassical state of
 the universe, which suggests a solution for the horizon problem
 different from that provided by inflation. }
 \received{???????? ?st, 2001} \accepted{???????? ?th, 2001}
 \preprint{\hepth{01??}\\RUNHETC-2001-5\\SCIPP-01/2\\UTTG-02-01}
 
\begin{document}
 %%%%%%%%%%%%%%%%%%%%%%%%%%%%%%%%%%%%%%%%%%%%%%%%%%%%%%%%%%%%%%%%%%%%%%%%%%%%
 % Table of contents automatic !!! %
 %%%%%%%%%%%%%%%%%%%%%%%%%%%%%%%%%%%%%%%%%%%%%%%%%%%%%%%%%%%%%%%%%%%%%%%%%%%%
 \section{\bf Introduction}
 
 In perturbative string theory, the only gauge invariant observable is
 the S-matrix. In modern times \cite{wibg}, this has been taken as
 evidence that the theory obeys the holographic principle \cite{ththsu}.
 In asymptotically flat spacetimes we can only probe it by sources
 localized on null infinity. In asymptotically Anti-DeSitter spacetimes,
 probes are similarly restricted to the boundary.
 The current technical machinery for understanding holography
 relies on a holographic screen at infinity. This is of course
 problematic when one tries to apply it to cosmology, particularly
 if we want to discuss closed universes. We have no doubt that the
 ultimate resolution of this puzzle will involve the recognition
 of a new gauge principle under which one can change the
 holographic screen without changing the physics. Indeed, the
 oldest holographic descriptions of M-theory vacua, Thorne's
 string bit models \cite{th} and Matrix Theory \cite{bfss} use a
 null hyperplane at infinity as the screen, via a choice of light
 cone gauge. It is clear that if the theory is Lorentz invariant
 we must be able to change the screen by a gauge transformation
 ({\it i.e.} pick a different light cone gauge). Much more
 definite evidence for the gauge nature of the choice of
 holographic screen comes from Bousso's work on entropy bounds in
 general spacetimes \cite{bousso}. There it was shown that the
 entropy (and thus the quantum states) of a general spacetime
 satisfying Einstein's equations and the dominant energy
 condition, could be assigned to a collection of holographic
 screens whose area counted the entropy. However, there is an
 enormous freedom in how the screens are chosen.
 
 It seems clear that a more local description of the theory will
 require us to obtain a deeper understanding of this new gauge
 invariance\footnote{L. Susskind has long advocated that a local
 description of M-theory would involve the introduction of many
 gauge degrees of freedom.}. One hint that was suggested in the
 talk by one of the authors (TB) at Strings at the Millenium
 \cite{millen} is a relation between this symmetry and local
 supersymmetry, via the twistor transform. We hope to return to a
 detailed discussion of this connection in a future publication
 \cite{susyfut}. For now suffice it to say that the choice of a
 (pure) spinor at a point P of spacetime or some brane embedded in
 it is equivalent to the choice of a null direction and an
 infinitesimal hyperplane transverse to it. This hyperplane should
 be thought of as the bit of the holographic screen on which the
 data at P is projected, and a gauge transformation which allows
 one to change the spinor is the required holographic gauge
 invariance. It is related to local SUSY in spacetime, and
 $\kappa$ symmetry on branes.
 The gauge variant nature of local physics is a generalization of the
 notorious {\it problem of time} in canonically quantized general
 relativity. It implies that many of the usual notions of local physics
 only make sense approximately, at low energies, and in regions where the
 spacetime curvature is small. What then are the mathematically well
 defined, gauge invariant observables of cosmological spacetimes? This is
 the question we will attempt to answer, for the case of asymptotically
 expanding universes, in the current paper.
 
 Morally speaking, our proposal can be defined in the classical
 approximation in terms of solutions of the field equations with
 boundary conditions on the Big Bang singularity and on a null
 surface in the future (null infinity if $\Lambda = 0$ and the
 future cosmological horizon(s) if the spacetime is Asymptotically
 DeSitter (AsDS)). There are two apparent problems with this
 description of the observables. The Big Bang is singular and
 the boundary value problem is not well defined there. In AsDS
 spaces, there are {\it many} cosmological horizons. The first of
 these problems will be solved when we have a complete quantum
 description of geometry. Some first tentative steps in this
 direction will be taken in this paper. The problem of multiple
 horizons in AsDS spaces will be seen to be related to the
 principle of Black Hole Complementarity \cite{thSUT}. This in
 turn is related to the {\it problem of time}: different observers
 in quantum spacetimes, have evolution operators that do not
 commute.
 
 It has been argued \cite{tbfolly} that AsDS spaces are described
 by quantum systems with a finite dimensional Hilbert space, with
 dimension equal to the exponential of the Hawking-DeSitter
 entropy. Bousso \cite{nbound} has recently proven that no
 experiment in a spacetime with positive cosmological constant can
 probe more information than this. We will argue that although
 there is no truly gauge independent physical S-matrix in AsDS
 spaces, there are gauge dependent objects which become invariant
 in the limit in which the cosmological constant vanishes. Close
 to this limit these objects represent physics as seen by low
 energy observers who have passed outside each other's horizon.
 The relation between them is similar to that between infalling
 and asymptotic observers for a very large black hole.
 We argue that the finite number of AsDS states is compatible with
 the indefinite growth of the number of horizon volumes in AsDS
 space because of a principle analogous to black hole
 complementarity. According to this principle, the Hilbert space
 describing the interior of a black hole is a tensor factor of the
 Hilbert space of the observer at infinity\footnote{All of this
 language is appropriate to time scales short compared to the
 black hole lifetime.}. The radical differences in the physics
 described by these observers is attributed to noncommutativity of
 the observables that they measure. Similarly, each observer in
 AsDS space sees everything in the space that is not bound to her,
 being absorbed in her cosmological horizon. Since the physics in
 these different horizon volumes is causally disconnected, there
 is nothing to stop us from using different bases of the same
 Hilbert space to describe all of them\footnote{This sentence is
 shorthand for the extensive discussion and explanation to be
 found in \cite{thSUT}.}. Thus, in an AsDS space, we will define an
 infinite number of different S-matrices, related by unitary
 transformations. As $\Lambda \rightarrow 0$ and the cosmological
 horizon recedes to infinity, these transformations approximately
 approach Poincare transformations relating observers at different
 points and in different Lorentz frames of an asymptotically flat
 space (though a real cosmology will never be exactly
 asymptotically flat and Poincare invariant).
 
 The rest of this paper is organized as follows. In the next
 section we formulate our prescription for the S-matrix in
 supergravity (SUGRA) language as well as for toroidal cosmologies
 of weakly coupled superstrings. We emphasize that all of these
 descriptions are inadequate because any cosmological solution
 contains a regime (which we call the Big Bang) where all weakly
 coupled or low energy descriptions fail. We give a preliminary
 semiclassical discussion of the physics of the Big Bang regime
 and show that it is dominated by matter satisfying the equation
 of state $p = \rho$, which is the equation of the homogeneous
 modes of moduli. This is the stiffest equation of state for
 which the speed of sound is less than or equal to that of light. 
 It is also the stiffest equation of state \cite{fs} for which the
 holographic principle can be satisfied at arbitrarily early times.
 We argue that moduli are not a satisfactory description of the
 Big Bang regime because they cannot saturate the entropy. A model
 which does saturate the entropy is one which postulates a $1+1$
 dimensional conformal field theory (whose central charge depends
 on the spacetime dimension) in each Planck scale volume of
 spacetime. The $L_0 + \bar{L_0}$ eigenvalue of the CFT is
 related to the initial energy density in Planck units. The CFT's
 in different Planck volumes are constrained to have the same
 eigenvalue of $L_0 + \bar{L_0}$.
 We also note that this model for the earliest state of the
 universe suggests a solution for the horizon problem different
 from that provided by inflation. This is implicit in the
 statement that homogeneous energy and entropy densities
 satisfying $p=\rho$ can saturate the maximal entropy allowed by
 the holographic principle. This means that a generic state of
 the universe near the Big Bang will have a homogeneous pressure,
 energy and entropy density in comoving coordinates, although its
 other characteristics need not be homogeneous ({\it i.e.} the
 state of the CFT in each Planck volume will have the same $L_0 +
 \bar{L_0}$ eigenvalue, but will not be the same state).

 The third section contains speculative material relevant for the
 construction of a true quantum theory of gravity. We argue that
 nets of finite dimensional Hilbert spaces (equivalently, finite
 dimensional $C^*$ algebras) can encode both the causal and metric
 properties of a spacetime satisfying the dominant energy
 condition and give a generalization of the notion of geometry to
 high curvature regimes that is based on nothing but the
 fundamentals of quantum mechanics. A Lorentzian manifold is
 determined by its causal structure, its conformal factor and its
 dimension. We suggest that near the Big Bang the dimension of
 spacetime is determined by the maximum rate at which information
 can be accumulated by an observer (this occurs for the $p = \rho$
 equation of state) and propose a way to match this to the
 behavior of nets of Hilbert spaces. The material in this section
 bears much resemblance to \footnote{but also, we believe, exhibits
 some differences from}, previous attempts to construct discrete
 theories of quantum gravity \cite{dqg}.
 In the fourth section we discuss the generalization of these
 considerations to AsDS spaces and explain why the natural
 substitute for boundary conditions on null infinity are boundary
 conditions at the cosmological horizon. We expand on the
 discussion of multiple horizons and complementarity given above.
 While this paper was being written we listened to E. Witten's talk
 at Strings 2001, using the magic of the Internet. There is some
 overlap with our considerations, and we devote section five to
 explaining our view of the relation between his talk and this
 paper.
 An appendix is devoted to a parable whose aim is to give
 philosophical solace to those who are disturbed by the notion of
 applying quantum mechanics to the universe.

 \section{\bf S-matrices in cosmology}
 
 \subsection{Classical Considerations}
 Let us approach the construction of S-matrices for cosmology from
 the point of view of classical field theory. Morally speaking,
 what we are looking for are solutions with boundary conditions on
 a Big Bang singularity as well as the future null infinity of an
 eternally expanding FRW cosmology\footnote{We will use the term
 FRW to refer not only to the standard homogeneous isotropic
 cosmologies, but also to anisotropic and possibly inhomogeneous
 cosmologies in which the spatial geometry is a compact Ricci flat
 manifold and may contain Wilson surfaces of p-form gauge fields.
 These are the most likely candidates for string/M theory examples
 of cosmologies with a future causal structure like that of the
 open and flat homogeneous cosmologies.}. This means that, in
 analogy with the solutions with Feynman boundary conditions in
 asymptotically flat spacetime, we are looking for {\it complex}
 solutions of the field equations. 

 However, the problem is much
 more involved than that of asymptotically flat spacetimes.
 First of all, we are asking for boundary conditions at a
 singularity, so that the mathematical problem is ill posed. We
 can finesse this for the moment by asking for boundary conditions
 a few Planck times (in a synchronous coordinate system) after the
 Big Bang. More importantly, most initial conditions at an initial
 value surface near the Big Bang, will not be compatible with a
 future evolution of the universe which corresponds to a few
 particles (infinitesimal wave packets) propagating in an
 asymptotically FRW universe. Indeed, we see immediately that the
 problem cannot be formulated in a purely classical manner. That
 is, the insistence on a final state consisting of well separated
 freely propagating stable particles might be viewed as ruling out
 solutions in which the final state contained large black holes.
 However we know that, in the fullness of time, and when quantum
 mechanics is taken into account, such states will indeed decay
 into a finite number of stable particles\footnote{Let us agree to
 work in more than four large dimensions, or simply to ignore the
 soft graviton infrared catastrophe.}. To take this into account
 at the classical level we have to accept final configurations
 consisting of any finite number of finite mass black holes plus
 any finite number of stable particles, propagating in an
 expanding FRW geometry. We cannot know what other kinds of
 restrictions and caveats we must put on a classical treatment of
 the S-matrix, without solving hard problems like the Cosmic
 Censorship Conjecture.

 The solution of this sort of boundary value problem is a highly
 nontrivial unsolved problem in classical general relativity. It
 is important to emphasize that it is very different from the
 initial value problem on a surface near the Big Bang
 singularity. Most initial conditions will not lead to an FRW
 universe containing only a finite number of particles and finite
 mass black holes in addition to the homogeneous background.
 Instead they will have singularities. If Cosmic Censorship is
 valid, then perhaps all solutions will evolve to black holes plus
 asymptotic particles in the future. We find it equally likely
 that some subclass of classical initial conditions will have to
 be rejected because they lead to physically unacceptable naked
 singularities \footnote{Something similar has been claimed
 recently about singular solutions in the AdS/CFT correspondence
 \cite{gubser}.}. Nonetheless, the set of solutions of the
 classical equations which {\it do} satisfy these boundary
 conditions will define a sort of classical approximation to the
 Cosmological Scattering Matrix. Note that the Scattering matrix
 defined in this way will be invariant under diffeomorphisms which
 vanish at the boundaries of spacetime.
 
Similar considerations can be applied to classical string theory.
 Consider a spacetime metric of the form \eqn{kasner}{ ds^2 = -
 dt^2 + R_i^2 (t) (dx^i )^2,} in weakly coupled Type IIA or IIB
 string theory. When supplemented by a time dependent dilaton
 field these solve the lowest order beta function equations if the
 $R_i (t)$ take the familiar power law, Kasner, form. They may
 be viewed as the dimensional reduction of the Kasner solutions
 for toroidally compactified 11 dimensional SUGRA. There exist
 solutions of these equations which at large positive times
 approach a slowly varying, infinite volume torus with weak string
 coupling. The results of \cite{bfmbm} show that the past
 asymptotics of these solutions is truly singular, in that there
 are no U-duality transformations which take them to weakly
 coupled string theories or to 11 dimensional SUGRA with slowly
 varying fields. Thus, in contrast to certain claims in the
 literature \cite{vatsbg}, string/M-theory appears to contain Big
 Bang cosmological singularities which cannot be removed by
 dualities. Despite the breakdown of string perturbation theory
 in these backgrounds, we can formally write down vertex operator
 correlation functions. At lowest order in $\alpha^{\prime}$ the
 vertex operators are in one to one correspondence with solutions
 with given asymptotics on the boundaries of the spacetime,
 including the Big Bang. Of course, since both the
 $\alpha^{\prime}$ and $g_S$ approximations break down near the
 Big Bang, we are again left without a true definition of the
 scattering matrix.

 \subsection{Semiclassical considerations}
 
Our finesse of the problem with the Big Bang Singularity was
 merely a stopgap measure. We believe that a true definition of a
 Cosmological S-matrix would have the following character: {\it
 The theory must simplify and become exactly soluble near the Big
 Bang, just as it does in the asymptotic future.}
 Then we could set up boundary conditions in terms of basis states of the
 Hilbert space which behave simply near the Big Bang. The
 Scattering matrix would be the amplitude to start in one of these
 simple basis states and end up in one of the simple states
 consisting of freely propagating FRW particles at null infinity.

 We do not have the required initial solution in hand, but believe
 a few clues to its nature can be obtained by thinking about the
 Fischler-Susskind holographic bounds \cite{fs} . A more
 fundamental approach to the initial singularity will be sketched
 in the next section. Fischler and Susskind argued that the
 holographic bound was compatible with FRW cosmology if the
 equation of state of the cosmic fluid was no stiffer than $p =
 \rho$. With this equation of state the holographic bound will
 remain saturated for all times if it is saturated initially. For
 softer equations of state, the homogeneous entropy density can
 only account for a fraction of the holographic entropy at late
 times. Note also that the $p = \rho$ equation of state is the
 stiffest for which the velocity of sound is less than or equal to the
 velocity of light. If we accept these two indications that $\rho
 = p$ is the stiffest equation of state allowed by nature, and
 further assume that there is some matter with this equation of
 state, then we can conclude that it dominates the earliest stages
 of the universe. Its energy density scales like the inverse of
 the square of the spatial volume\footnote{We note, without claiming 
 to understand the connection, that 't Hooft has used the $p=\rho$ 
 equation of state in a model of black hole entropy\cite{tbh}.}.
 
 What can this primordial form of matter be? One possibility is
 the homogeneous modes of various massless fields. This includes
 massless minimally coupled scalars, topological modes of p-form
 gauge fields, and unimodular deformations of the space of
 Ricci-flat spatial metrics on a manifold with compact spatial
 sections. We will refer to these collectively as moduli.
 A problem with moduli from the point of view of the holographic
 principle is that they do not provide an entropy density but
 rather a finite entropy for the total spacetime. This would not
 appear to be a significant difference for a spacetime with closed
 spatial sections. The Fischler-Susskind description refers to a
 comoving entropy density, which, if the spatial manifold is
 closed, leads to a finite total entropy.
 
 The important deficiency of moduli is revealed when one attempts
 to saturate the holographic bound. This leads to an equation of
 the form $\sigma_0 \propto \sqrt{\rho_0}$, where $\rho_0$ is the
 initial energy density of the system, and $\sigma_0$ its
 (constant) comoving entropy density. For moduli, $\rho_0$ is
 the initial energy of the homogeneous modes, divided by the
 volume of comoving coordinate space. It is easy to see that the
 entropy of moduli varies logarithmically with $\rho_0$. Instead,
 the formula suggests that in each Planck scale cell of comoving
 coordinate space there is a conformally invariant $1+1$
 dimensional field theory (the CFT at the beginning of the
 universe). The states of these CFT's are constrained to have the
 same $L_0 + \bar{L_0}$ eigenvalue, but are otherwise independent
 of each other.
 It is not immediately apparent that such a CFT satisfies the
 equation of state $p = \rho$ in {\it spacetime}, since the
 mapping between its world volume and spacetime has not been
 specified. However, there is a general thermodynamic argument
 for homogeneous systems that the relation $\sigma \sim
 \rho^{1/2}$ implies the $p=\rho$ equation of state.
 In each comoving volume, the energy density, entropy density,
 temperature and pressure are related by
 \eqn{loceq}{\rho = T \sigma - p .}
 This equation follows from the laws of thermodynamics
 (\Ref{thermo}) if we assume the energy and entropy densities, and
 pressure, are independent of the volume. Combining \Ref{loceq}
 with the general thermodynamic equation
 \eqn{thermo}{dE = T dS - p dV,}
 applied to variations that leave the volume unchanged, and with
 the relation $\sigma = \alpha \rho^{1/2}$, we obtain
 \eqn{temp}{T = {2\over\alpha} \rho^{1/2}.} Returning to
 \Ref{loceq} we find that $p=\rho$. Thus a model in which each
 comoving Planck volume has associated with it a $1+1$ dimensional CFT
 fits the
 phenomenology of a holographic universe and can saturate the
 holographic bound.

 In fact, the requirement of saturation allows us to calculate the
 central charge of the conformal field theory in each Planck cell,
 if we normalize the length of the interval it lives on to $2\pi$.
 Saturation of the holographic entropy bound requires that the
 entropy and energy in Planck units in a unit Planck volume, are
 related by $\sigma_1 = {1\over 4} (d-2)\sqrt{\rho_1}$.In the
 CFT the relation is $\sigma_1 =  \sqrt{ c \rho_1 /6}$, where
 $\rho_1$ is the eigenvalue of $L_0 + \bar{L_0}$ .  The two
 formulae are identical if $ c = {3\over 8} (d-2)^2 $.
 We do not yet understand the significance of this result, which
 one would have hoped to be a clue to the nature of the CFT.

 As noted in the introduction, our constraint that different cells
 have the same energy density is a solution of the horizon problem
 of cosmology. We note that this is not just putting in the
 answer by hand. Our considerations show that a homogeneous
 system can saturate the maximal entropy allowed by the
 holographic bound. This tells us that the initial conditions for
 the universe are much more highly constrained than local field
 theory would have led us to believe. In local field theory one
 can find homogeneous solutions of the equations of motion, but
 there appear to be a host of inhomogeneous perturbations of them.
 The holographic principle restricts the allowed perturbations
 because it bounds the entropy in regions of spacetimes which are
 close to the homogeneous cosmology. Furthermore, if $p = \rho$,
 and the initial entropy and energy density are related in the
 proper manner (corresponding to the correct choice of central
 charge in the CFT model) then we have a homogeneous system that
 saturates the entropy bound. Thus there cannot be any nearby
 states which are inhomogeneous. Our decision to impose the
 constraint of equal energy for the CFT in each Planck cell was an
 attempt to model this fact about the holographic bounds, rather
 than to solve the horizon problem in an {\it ad hoc} manner.

 We emphasize that this would be solution for the horizon problem
 will only make sense if we find a complete theory from which we
 can derive the phenomenological CFT model presented in this
 section. Further, there is no sense at the moment in which our
 considerations can be viewed as a replacement for inflationary
 cosmology. We do not have any evidence that our model solves any
 of the other cosmological conundra that are traditionally cited as
 evidence for inflation. Most importantly, we do not yet see a
 viable alternative model for the fluctuations which produce
 galaxies.
 In the next section, we will attempt to formulate a more basic
 approach to the dynamics of the Big Bang. Eventually, one would
 hope to derive the CFT model presented here from such a
 fundamental approach.

 \section{\bf Prolegomena to a quantum theory of space-time}

 In this section we wish to present the beginning of an attempt to
 formulate Quantum Cosmology in a fundamental fashion. The
 formalism we will present is incomplete. It was discussed
 previously in the talk of one of the authors (TB) at the Strings
 at the Millennium Conference. The basic observation is that when
 the Fischler-Susskind-Bousso (FSB) bounds are applied to regions
 of spacetime consisting of the intersection of the causal pasts
 of a finite number of points, they give a holographic derivation
 of the concept of {\it particle horizon} in an expanding Big Bang
 universe. Since the notion of particle horizon is usually
 derived from locality, this is a clue to the relation between
 locality and holography. That relation has remained completely
 obscure in the Matrix Theory or AdS/CFT versions of holography.
 We argue that if we take the FSB entropy to be an actual count of
 the number of states, rather than just a bound, a picture of the
 Universe emerges in terms of a net of interlocking operator
 algebras. What is remarkable about this picture is that
 geometrical facts about a Lorentzian spacetime are translated into
 algebraic facts about quantum operator algebras. We argue that
 the quantum formalism can encode both the causal and metrical
 structure of the spacetime, and conjecture that (once a full set
 of axioms for the nets of algebras is discovered) there will, in
 the limit of large dimension algebras, be a unique spacetime
 corresponding to a given net of algebras.

 A Lorentzian spacetime is characterized by its dimension, its
 causal structure and its conformal factor. The axioms we have
 formulated so far give information about the last two kinds of
 geometrical data, but not about the dimension. We sketch a
 program for understanding the spacetime dimension as well as the
 axioms we believe are missing. The idea is to find the algebraic
 structure corresponding to the $p = \rho$ state of the universe
 that we discussed above. Since this appears to characterize a
 situation in which, for a given spacetime dimension, we have the
 fastest rate of growth of information inside the horizon, we can
 hope to discover it by finding the fastest consistent growth rate
 of information in the algebraic formalism.
 Unfortunately, it is precisely the rules for consistent time
 evolution that we do not understand in the algebraic formalism.
 Thus, our discussion is just a sketch of a plan for discovering a
 fully consistent theory of quantum cosmology.

 Consider a spacetime that begins with a Big Bang Singularity.
 That is, there is a spacelike surface on which the universe
 begins. Furthermore, the proper time between any point in the
 spacetime and the Big Bang surface is finite. Let us define a Past
 Intersection Region (PIRE) as the intersection of the causal pasts
 of a finite number of points. A PIRE which is the causal past of
 a single point will be called a basic PIRE. The boundary of a PIRE
 is an almost everywhere null region. Thus, there is an FSB bound
 for the entropy flowing through the boundary of any PIRE. This
 can be viewed as the maximum entropy that could be observed in any
 experiment done inside the PIRE. We will assume that this entropy
 bound is in fact saturated for the basic PIREs. R. Bousso has
 suggested to us that this cannot be so for generic PIREs, whose
 light sheets are artificially truncated by the surfaces of
 intersection. That is, we can associate to each basic PIRE a
 Hilbert space whose dimension is given by the exponential of the
 area of a holographic screen for the PIRE. Notice that all of
 these dimensions are finite, as a consequence of the finite
 proper time to the Big Bang. For generic PIREs we construct a
 Hilbert space whose dimension is the exponential of the area of
 the largest light cone that fits inside the PIRE.
 
 Consider now some region of an expanding universe and two PIREs
 $P_1 \subset P_2$. It follows immediately from the holographic
 bounds that an observer in $P_1$ can observe fewer states than an
 observer in $P_2$ (equivalently, his observables constitute a
 smaller operator algebra) . This only makes sense if in fact the
 operator algebra $A_1$ is a tensor factor in $A_2$ {\it i.e.}
 $A_2 = A_1 \otimes \bar{A}_{12}$. Otherwise, observations in
 $P_1$ would not commute with observations in the part of $P_2$
 disjoint from it, and there would no way to consistently discuss
 the {\it information in $P_1$}. The fact that a PIRE has an
 operator algebra that commutes with the algebras of other regions
 is usually derived from locality. In local field theory one
 assumes that every spacetime region has an operator algebra
 associated with it that commutes with the algebras of all other
 regions that are separated from it by a spacelike interval. This
 principle of locality is not compatible with the holographic
 principle.
 
 However, we see that the holographic principle itself forces us
 to the concept of {\it particle horizon}, which we usually derive
 from locality. Each PIRE has only a finite number of states
 associated with it and smaller PIREs have a smaller number of
 states. One can dimly glimpse how our conventional concepts of
 local evolution may be compatible with holography.
 
 Thus, we propose to associate a finite dimensional Hilbert space
 to every PIRE. Furthermore, every pair of PIREs has an
 intersection (possibly empty) which is also a PIRE. Thus if
 $H_I$ and $H_J$ are the Hilbert spaces of two PIREs then we must
 have
 \eqn{PIRINT}{H_I = H_{IJ} \otimes D_{IJ},} \eqn{PIRINTb}{H_J =
 H_{JI} \otimes D_{JI},} \eqn{PIRINTc}{U_{JI}^{-1} = U_{IJ}: H_{IJ}
 \rightarrow H_{JI}.} That is, each Hilbert space must have a
 tensor factor which is shared between the two and represents the
 Hilbert space of the intersection. The invertible unitary
 isomorphism $U_{IJ}$ takes into account the possibility of a
 different choice of basis made by observers in each PIRE for this
 common Hilbert space. The relations between members of this
 collection of Hilbert spaces encode information about the causal
 structure of the spacetime, and their dimensions tell us
 something about its geometry.
 
 Now let us consider a network of basic PIREs, the tips of whose
 light cones are only a Planck distance apart along spacelike or
 timelike geodesics. Note that such a network inevitably involves
 some coordinate choices: there are many inequivalent ways of
 choosing a lattice of Planck separated points in a spacetime. We
 call any such choice, a {\it Planck lattice}. It seems clear that
 the net of PIREs associated with a Planck lattice, along with the
 areas of their holographic screens, should determine the geometry
 of spacetime, with at least Planck scale accuracy. Since the
 information about the PIREs, their overlaps, and their
 holographic areas, can all be encoded into properties of a net of
 Hilbert spaces, we will propose that quantum spacetimes are
 simply such a net, obeying appropriate axioms.
 
 One final point before proceeding to this program. If we have two
 PIREs such that $P_1 \subset P_2$ then it is clear that $H_1 =
 H_{12}$. However, the converse is not necessarily true. Consider
 for example an FRW universe with vanishing cosmological constant
 and a black hole embedded in it. If we examine the causal past of
 a point inside the horizon of the black hole, the holographic
 screen of this PIRE lies entirely outside the horizon and can be
 completely contained in the causal past of points in the
 asymptotic region. In this case we would also say $H_{BH} =
 H_{BH,\infty}$ even though part of the black hole PIRE is
 causally disconnected from the outside. The inclusion relations
 between Hilbert spaces reflect those between holographic screens,
 rather than spacetime regions. This rule is necessary in order
 to construct a formalism, which does not have a black hole
 information paradox.

 \subsection{Geometry from quantum mechanics}
 
 Now we would like to turn these relations around and propose a
 set of axioms for nets of Hilbert spaces\footnote{Equivalently,
 since we are dealing with a finite dimensional situation, for
 nets of operator algebras.} that will allow us to reconstruct a
 spacetime. We will be only partially successful in this endeavor.
 We begin with a list of a countably infinite set of finite
 dimensional Hilbert spaces $H_n$. Each of these are supposed to
 represent the quantum observables in the causal past of a single
 point. We emphasize however that the quantum mechanics is
 fundamental, while the geometrical interpretation of it is only
 supposed to emerge in a limit of very large dimension Hilbert
 spaces. For each pair of Hilbert spaces we have a set of
 equations analogous to \Ref{PIRINT}:
 \eqn{PIRINT2}{H_m = H_{mn} \otimes D_{mn},} \eqn{PIRINT2b}{H_n =
 H_{nm} \otimes D_{nm},} \eqn{PIRINT2c}{U_{nm}^{-1} = U_{mn}:
 H_{mn} \rightarrow H_{nm}.}
 Now we add the $H_{mn}$ to our list and repeat the procedure.
 However, we must also add rules which assure that {\it e.g.}
 $H_{mn,k} = H_{m,nk} = H_{k,nm}$ {\it etc.}. All of these spaces
 must have the same dimension and there must be unitary mappings
 between them which are compatible with the unitary embeddings of
 single overlaps into their parent $H_i$. Thus there will be a
 single Hilbert space $H_{ijk}$, symmetric in all indices, which
 can be embedded as a tensor factor of each Hilbert space with one
 or two of the indices $(i,j,k)$. In a similar manner we can
 build $k$ fold overlaps $H_{i_1 \ldots i_k}$. We will consider a
 set of multiply indexed Hilbert spaces satisfying the above
 tensor inclusion relations (plus some other axioms that we do not
 yet understand) to be the quantum version of a cosmological
 spacetime.

 We view the above construction as the analog of constructing an
 atlas of charts which covers a Lorentzian manifold. These charts
 contain information both about causal structure and metrical
 geometry. The former is encoded in the inclusion relations
 between Hilbert spaces, which should be mapped into the causal
 relations between holographic screens. The dimensions of the
 Hilbert spaces encode metrical information. One would like to
 prove that in the limit of large
 dimensions, such a Hilbert space construction in fact determined
 a unique Lorentzian spacetime.
 We believe that there are a number of axioms which must be added
 to our system before one can hope to prove such a theorem. We
 will describe a number of different problems that we do not
 know how to solve, but they may not  all be independent.

 The first set of problems has to do with the analogy to the
 definition of a manifold. There one starts from an abstractly
 defined topological space that satisfies the Hausdorff property
 that any two points can be separated by open sets. Our
 holographic geometries do not have points and it is not clear to
 us what the analog of the Hausdorff property is. Furthermore,
 the abstract topological space in the definition of a manifold
 allows us to define the idea that an atlas covers the manifold.
 In what we have done so far the charts themselves seem to define
 the manifold. We believe that this is a symptom of a very deep
 property of the holographic approach to spacetime: one cannot
 decide whether two regions of spacetime are independent of each
 other (have independent degrees of freedom) until one knows their
 ultimate fate. Thus, a lengthy inflationary period followed by
 reheating and a matter or radiation dominated FRW cosmology,
 looks like an AsDS space for many e-foldings. If the DS era
 lasts forever, so that the cosmological horizon is a true
 horizon, then we consider the degrees of freedom in different
 horizon volumes to be gauge copies of each other. The experience
 of different causally disconnected observers is thought of as
 arising from a different choice of basis in the same Hilbert
 space. On the other hand, in the inflationary cosmology we
 consider these degrees of freedom as independent (and use them as
 a basis for calculating cosmic microwave background fluctuations)
 because an observer in the far future of the inflationary era will
 be able to measure correlations between them.

 Another example that illustrates the same problem can be
 constructed by thinking about a Big Bang Universe which
 asymptotes to DS space. As in the previous section, and
 according to standard convention, one describes the early
 universe in comoving coordinates and assigns it a homogeneous
 entropy density. If the equation of state at early times is $p =
 \rho$ then such a picture can capture the correct holographic
 counting of entropy at arbitrarily early times, bounded from
 below only by our decision about when the semiclassical picture
 breaks down. Furthermore, assuming compact spatial sections,
 there is nothing in the FSB discussion that restricts the
 comoving coordinate volume. Thus, we could assign such a
 universe an arbitrarily large total entropy, using conventional
 language. On the other hand, one can construct models in which
 such a universe asymptotes to a DS space with DS entropy less
 than the total entropy of the universe at early times. There is
 no contradiction here because this simply means that many
 different coordinate volumes will be outside each other's
 particle horizon forever. The holographic principle again tells
 us to treat these degrees of freedom in terms of different bases
 in the same Hilbert space, rather than independent Hilbert
 spaces. But there is no way to make this judgement without
 knowing the future asymptotics of the cosmology. This makes
 sense when thinking about covering a Lorentzian manifold by
 PIREs, but is at odds with the usual notion of physics being
 determined by initial conditions. It is more compatible with the
 Feynman propagator or S-matrix approach, in which boundary
 conditions on both past and future are necessary to completely
 specify a physical process.

 This brings us to the most serious difficulty with our formalism,
 which is how to construct a unitary time evolution operator. We
 should recognize from the beginning that their should be no
 unique prescription for such an operator. Different physical
 coordinate systems should have different evolution operators. In
 the semiclassical approach to quantum gravity, time evolution is
 defined in terms of certain approximately classical
 variables\cite{rtbfs}. These can be viewed as defining a
 classical background geometry. Even in this situation there can
 be various natural definitions of time which are not related by
 isometries.

 Quantum mechanics is usually described in terms of a Hilbert
 space and a Hamiltonian operator. It is important to recognize
 that the Hilbert space tells us almost nothing about the system.
 Any two separable Hilbert spaces of the same dimension are
 unitarily isomorphic. It is the Hamiltonian that defines what we
 usually think of as ''the structure of the Hilbert space of the
 system". For example, many Hamiltonians have the property that
 their asymptotic high energy spectra are identical to those of a
 system of some number of Gaussian variables. One can describe
 any Hamiltonian with the same Gaussian fixed point in terms of
 differential operators acting on wave function(al)s of the
 Gaussian variables. This is the conventional formalism of
 Schrodinger quantum mechanics.
 In a system without an {\it a priori} Hamiltonian one might
 imagine {\it any} one parameter group of unitaries (or perhaps
 just a discrete unitary group if we give up the idea of
 continuous time evolution- see below) be viewed as time
 evolution for some observer. It is only in a regime
 where the system has many states and enough variables to behave
 classically that one could dismiss many of these would be
 observers as ''unphysical". Unphysical evolutions would be
 defined to be those which do not preserve the classical nature of
 the classical variables.

 In our formalism, the Hilbert space itself has much more
 structure. The whole system can be mapped into a single Hilbert
 space, the space of the asymptotic future\footnote{We restrict
 attention to cosmologies that expand indefinitely. The Big
 Crunch requires a separate discussion.}. This is defined as a
 Hilbert space $H_{\infty}$ such that any of the $H_n$ are tensor
 factors of it. Furthermore, we insist (as an additional axiom if
 necessary) that there is some sequence of $H_{n_i}$ which
 converges to $H_n$. However, any definition of an evolution
 operator on $H_{\infty}$ must be compatible with the causal
 structure that is implicitly defined by the $H_{n_1 \ldots n_k}$.
 Let us examine some of the consistency conditions that a time
 evolution must satisfy. First, it must ''foliate" the
 collection of Hilbert spaces , $H_{n_1 \ldots n_k}$. That is, it
 must break the collection up into subsets, each of which is ''on
 a given time slice". We denote this by giving each Hilbert space
 a label $t$. Thus, the indices $n$ of the previous discussion
 are broken into a composite index $(t,n)$, where we abuse the
 reader's patience by using the same letter for the ''spatial"
 part of the composite index as we previously used for the whole.
 The slicing must be compatible with the existing causal structure.
 Thus, every $H^{t_1 \ldots t_k}_{n_1 \ldots n_k}$ with all $t_i
 \leq t$ should be a tensor factor of some $H^{t+1}_n$.
 The real problem is to find a map which gives us a unique state
 in each of the Hilbert spaces whose largest time index is $t+1$
 given a state in each of the Hilbert spaces with largest index
 $t$. It is plausible that if it is possible to find time slices
 for which such a map exists, that there will be many inequivalent
 ways to do so. This would correspond to the many fingered time
 of general relativity. However, it is clear to us neither
 whether it is always possible to find such a time slicing, nor
 what axioms have to be added to the structure of our net of
 Hilbert spaces in order to guarantee that such a slicing exists.
 Another puzzle is the question of the uniqueness of the map
 between successive slices and what the proper interpretation
 would be for finding many consistent maps. At the present we do
 not have an answer to any of these questions, which explains the
 title of this section.

 The reader will have noted that the time evolution we describe is
 discrete. This seems to follow inevitably from the idea that
 time evolution always follows expansion of the size of the
 particle horizon plus the fact that area is quantized in bits. We
 should identify the time step with the Planck size time steps in
 the lattice of basic PIREs described in the previous section.
 Indeed, it seems that the right way to understand the problem of
 time evolution is to try to construct nets of Hilbert spaces
 which satisfy the inclusion relations of the Planck spaced web of
 PIREs that we constructed in the previous section. We will
 reserve this project for a future paper, apart from the remark
 that this construction makes it clear that our nets of Hilbert
 spaces represent a {\it gauge fixed} construction of the theory.
 We could describe the same spacetime by constructing the net of
 Hilbert spaces of many different Planck lattices. These are all
 {\it physical} gauges , in the sense that all of the Hilbert
 spaces have positive definite metric. They should represent the
 physical measurements of observers who choose different
 coordinates to describe spacetime (and have also made choices of
 holographic screen that are locked to the coordinate choice),
 whenever the evolution is sufficiently classical to justify the
 separation of the system into an observer and the rest of the
 universe. Nonetheless, none of this information is strictly
 gauge invariant. We reserve to the future the important task of
 formulating the equivalence relation between nets of Hilbert
 spaces implied by general coordinate invariance, as well as the
 additional axioms that will guarantee that a net of Hilbert
 spaces can be realized as the net of PIREs associated with a
 Planck lattice in a spacetime satisfying the dominant energy
 condition.

 Another very interesting question, about which we have only
 conjectures, is the extent to which information about the universe
 is encoded purely in the net of Hilbert spaces rather than the
 choice of state in these Hilbert spaces. One might imagine that
 consistency conditions of some sort were strong enough to
 completely specify the state \footnote{We owe this radical
 conjecture to R. Bousso.}. We believe that it is more likely
 that in the semiclassical limit there will be a sense in which
 the net of Hilbert spaces contains information only about the
 geometry, while the state will tell us about the properties of
 branes propagating in the geometry. As a consequence of
 M-theory dualities, even this cannot be exactly right under all
 circumstances. But such ambiguities really only arise when the
 universe has small dimensions and neither of the U-dual notions of
 geometry is valid. As all dimensions become large, the
 separation between geometry and branes becomes sharp. The
 property that not all the information in our systems is in the
 structure of the net of Hilbert spaces is shared by our $1+1$ CFT
 model of the early universe. There, the geometry is determined
 by and determines only the $L_0 + \bar{L_0}$ eigenvalue of the
 CFT, but not its state.

 Finally, we can return to our discussion of the S-matrix. If our
 ideas about Planck lattices are correct, it will be very easy to
 impose the condition that the net of Hilbert spaces
 asymptotically approaches that of an eternally expanding FRW
 universe with a boundary causally equivalent to that of Minkowski
 space. The advantage of our present formalism is that we can make
 coherent remarks about the initial state. Indeed, it is obvious
 that as we go back in time, the particle horizon becomes smaller
 and smaller. At some point, its area in Planck units becomes of
 order $4 {\rm ln} 2$. The dimension of any Hilbert space on this
 slice of comoving coordinates is no bigger than two. We cannot
 extrapolate the evolution back any further. This is the point
 we wish to identify with the Big Bang. The quantum state of the
 universe is a tensor product of independent states in a
 collection of two state systems. There are no more overlaps,
 because 2 is a prime number\footnote{D.Gross and L.Motl pointed out
 to us that any other prime might do as well, and that one could
 imagine different primes in different initial Hilbert spaces.
 This issue will have to be confronted if we are to make believable
 claims about noninflationary solutions to the horizon problem.}. 
 In an expanding universe with
 vanishing cosmological constant the number of two state systems
 will be infinite. 

 Note that in making these statements we have
 not had to specify any particular properties of the net of
 Hilbert spaces. Thus this initial state will be gauge invariant
 under the equivalences that we have discussed above. The
 detailed properties of the spacetime are encoded in the evolution
 rules which generate the rest of the net. These, as we have
 discussed, are gauge variant. By imposing asymptotic conditions
 on the net in the infinite future, we obtain a set of completely
 gauge invariant amplitudes. The different but equivalent nets
 will represent different asymptotic coordinatizations of the
 system, but, in the conventional manner, these are not imposed as
 gauge symmetries\footnote{Usually we only discuss the asymptotic
 isometries of {\it e.g.} asymptotically flat spaces, but the
 whole set of coordinate transformations that act nontrivially on
 the asymptotic space can be viewed as physical operations on the
 Hilbert space of the system.}. If the eventual fate of the
 universe were indeed an FRW spacetime, we could choose to restrict
 our attention to nets that approached those defined by Planck
 lattices in comoving coordinates.

 There is a further curious possibility that our ignorance of the
 full set of axioms prevents us from making a definitive statement
 about. Recall that the embeddings of earlier Hilbert spaces as
 tensor factors of later ones involve unitary mappings. Then it
 may be that the choice of initial tensor product state is a gauge
 choice, {\it i.e.} that the initial state is unique up to gauge
 transformations. If this is the case, then all information about
 which universe one was describing, would be encoded in the laws
 of time evolution.  This would mean that the problem of time
 would be mixed up with what is generally called the question
 of initial conditions.

 Thus, we believe that our formalism will be able to give a
 completely well defined and gauge invariant definition of an
 S-matrix for FRW spacetimes with vanishing $\Lambda$. The
 initial state is a tensor product state of an infinite collection
 of two state systems. The final state consists of a finite
 number of stable particles propagating on an FRW spacetime.
 The tensor product state of two state systems does not bear any
 resemblance to a spacetime. The conventional semiclassical
 regime will set in only after propagation to the point where the
 Hilbert spaces in our net define areas large compared to the
 Planck scale. It is in the early stages of the semiclassical era
 that one can expect the $p = \rho$ phase to occur. It seems to
 be the simplest semiclassical cosmological era and we hope to
 encode its properties in the rules for evolving the net. This
 project as well will be reserved for a future paper.
 
To summarize this meandering section: It is clear that the set of
 PIRES (and their holographic areas) generated by a Planck lattice
 contains enough information to determine the spacetime in which
 they are embedded. Since all this information can be translated ,
 via the holographic principle, into information about a net of
 interlocking Hilbert spaces (operator algebras), we have a strong
 indication that a set of axioms formulated solely in terms of such
 Hilbert nets can reproduce classical general relativity as an
 approximation. We have formulated some, but (we are quite sure)
 not all, of the axioms. The formalism promises to give a
 picture of black hole formation and decay which is manifestly
 free of paradoxes, and incorporates the Black Hole Complementarity
 principle. It gives a picture of the Big Bang as a collection of
 decoupled two state systems. The gauge invariances of the
 formalism may imply that there is a unique initial state of this
 system. We hope that by studying the transition of this system
 into a quantum model of $p = \rho$ cosmology, we will be able to
 learn the rules for constructing a general quantum geometry.
 
 \section{\bf Gauge variant S-matrices for AsDS spacetime}

 We now want to generalize our considerations to the case of AsDS
 spacetimes which begin with a Big Bang singularity. In our
 opinion, the proper generalization is a matrix which interpolates
 between states at the Big Bang (which we now recognize as tensor
 product states in a collection of two state systems) and states on
 the cosmological horizon. As we take $\Lambda$ to zero, a single
 horizon volume in DS space approaches all of Minkowski space,
 with the horizon approaching the boundary of Minkowski space. A
 similar statement is valid for AsDS cosmologies and FRW
 cosmologies with vanishing $\Lambda$. What is confusing, is that
 there are an infinite number of different cosmological horizons
 in AsDS space. So our prescription has an ambiguity.
 Furthermore, the different horizon volumes are mapped into each
 other by diffeomorphisms so no given S-matrix is gauge invariant.
 
 We believe that this correctly represents fundamental physical
 properties of AsDS spaces\footnote{which L. Susskind has
 characterized as ''the great crisis in theoretical physics that
 would be caused by the observational proof that there is a
 positive cosmological constant". As will be clear below, we take a
 somewhat more sanguine view of the situation.}. The key to
 understanding this is the Black Hole Complementarity Principle of
 't Hooft, Susskind, Thorlacius and Uglum \cite{thSUT}, and we will
 begin by briefly reviewing these arguments.
 Let us remind the reader that although a plausible case has been
 made that scattering off a black hole is unitary, we have as yet
 no clues about how to describe the experience of the infalling
 observer. The BHCP is a slogan that outlines what such a
 description might look like. We will summarize the arguments of
 \cite{thSUT} by the statement that thought experiments show that
 no comparison between the states of the infalling and external
 observers in a black hole is possible within the realm of low
 energy effective field theory (which is the only realm in which
 the concept of infalling observer is clearly defined). HSUT argue
 that this means that the claim that the external observer's
 Hilbert space of scattering states is complete is not ruled out.
 In such a description, the infalling observer makes measurements
 in the same Hilbert space, but measures observables which are
 complementary to those of the external observer. One cannot
 compare the measurements because they interfere with each other.
 Since the infalling observer's time evolution is finite, the only
 precisely defined gauge invariant observable of the system is the
 external S-matrix. The quantum mechanical definition of the
 infalling observer is neither precise nor gauge invariant, except
 perhaps in the formal limit of infinite black hole mass.
 
 Now consider the situation in an AsDS space. For any given
 observer, all things that are not bound to her appear to be
 squeezed closer and closer to her horizon volume as the Hubble
 flow sweeps them away. All that remains of them is a thermal gas
 of Hawking radiation. If the observer is part of a large,
 gravitationally bound system, then she is likely to eventually end
 up as an infalling observer for a large black hole. After the
 black hole decays, its remnants are swept into the Hawking gas
 clumped near the DS horizon. All that remains is a complementary
 (but decidedly uncomplimentary) image of the Hilbert space states
 that once described her existence. If she has cleverly avoided
 this fate by building a large steel living module somewhere in the
 space between galactic clusters, her agony is only prolonged. The
 Second Law of Thermodynamics assures us that she will eventually
 be thermalized and become part of the Hawking DeSitter gas.
 Thus, {\it all} observers in AsDS space, regardless of sex, race
 creed or color, are analogous to infalling observers for a large
 black hole. The true final state of the system is always one of
 the states of the thermal ensemble. 

 The thermal ensemble is DS
 invariant and therefore gauge invariant. Furthermore, once we
 agree that, asymptotically, there are no macroscopic objects left
 in DS space, one might imagine that the only sense in which
 individual states that make up this ensemble fail to be gauge
 invariant is that they are mapped into identical states in another
 horizon volume. These would then have a gauge invariant
 definition as well.
 The question of the nature of these states depends
 on details of the full theory of quantum gravity that we do not
 yet understand.   The microscopic theory of these states might
 show that the natural basis for describing them was one in which
 the observables describing finite macroscopic objects localized
 in DS space were not diagonal. One could then imagine constructing
 identical bases in the Hilbert spaces of different horizon volumes
 and thus a completely gauge invariant S-matrix for AsDS systems.
 But if the S-matrices for different horizon
 volumes were identical they could contain no information about the
 existence of macroscopic observers with different experiences in
 different horizon volumes. So, either there is no gauge invariant
 S-matrix, or it does not contain information about the physics
 that we think is interesting.

 The difficulties inherent in describing gauge invariant DS physics,
 are closely connected with the {\it problem of time} of traditional 
 quantum cosmology. There are no preferred
 unitary operators in a cosmological Hilbert space except in
 extreme circumstances where semiclassical approximations are
 valid\footnote{and, if our conjectures are correct, very near the
 Big Bang singularity, where we hope that the physics becomes
 soluble, though not classical.}. We have already remarked above
 that {\it all Hilbert spaces look the same in the dark}. Without
 the guide provided by a Hamiltonian, we cannot distinguish
 different systems with the same number of states. Thus, without a
 compelling set of semiclassical observables, there is no reason
 to prefer one sequence of unitary transformations in Hilbert
 space from another. Our causal nets of Hilbert spaces provide
 us with a little more structure, but at least to some extent this
 corresponds to a choice of coordinates. AsDS space and black hole
 physics present us with an even more disturbing situation in
 which there may be several different semiclassical descriptions,
 that are complementary in the sense that their time evolution
 operators do not commute with each other (this would be the
 mathematical statement of the BHCP). In the case of black holes,
 the external observer's semiclassical description is preferred,
 because it is completely well defined, while the infalling
 observer has only a finite lifetime and cannot be expected to
 have a complete and exact description of physics. In AsDS space,
 all macroscopic observers are equal, and equally ill defined.

 Our interpretation of the apparent lack of gauge invariance of
 observables in AsDS spaces can thus be phrased as a DS
 Complementarity Principle. More generally, we interpret the
 Problem of Time as a Cosmological Complementarity Principle. That
 is, different physical observers in a Cosmological spacetime will
 generically have time evolution operators that do not commute with
 each other. The Complementarity principle tells us that these
 operators all act in the same Hilbert space, even when we are
 referring to two sets of observations that are out of causal
 contact forever. In some cases, there will only be a single set
 of semiclassical observables and we can choose these to define
 special classes of gauge transformations under which the physical
 Hilbert space is not required to be gauge invariant. The
 asymptotic observables in various kinds of asymptotically infinite
 spacetimes are a particular example. AsDS spaces are different
 because there are in principle an infinite number of equally good
 semiclassical time evolution operators at late times.

 The difficulties of interpretation of AsDS physics are also
 connected to the finite dimension of the Hilbert space that
 represents AsDS space\cite{tbfolly}. We suspect that for Hilbert
 spaces with dimension $2^N$ with $N$ which is not enormously
 large, there will be only a few causal nets of Hilbert spaces that
 can be constructed. None will have a spacetime interpretation and
 no consistent unitary evolution will be preferred over any other.
 There is unlikely to be gauge invariant physics associated with
 such a system. In a universe with only a few states, a basic
 assumption of all physical theories\footnote{not just quantum
 mechanics}, becomes untenable. This is the claim that one can
 separate the system into {\it observer} and {\it observed} with
 sufficiently small interaction between them that one can make a
 measurement without completely changing the property one was
 trying to measure, or destroying the measuring apparatus.
 As $N$ becomes large the possibility of forming large
 classical subsystems emerges. We have argued that all such
 subsystems have finite lifetime. However, we should remember that
 if our universe has a cosmological constant, the value of $N$ is
 $10^{123}$ . The finite lifetime for classical subsystems is of
 order the infall time for all galaxies to collapse completely into
 black holes (for gravitationally bound systems) or of order the
 thermalization time (for isolated space platforms). We note that
 the average number of Hawking particles in the Hawking-DeSitter
 gas is one per horizon volume\footnote{These most probable
 components of the radiation also have wavelengths of order the horizon
 volume and have little effect on localized objects.  The probability of
 finding dangerous shorter wavelength radiation is even smaller.}
 , so the thermalization time is
 unimaginably long. Thus we believe that for very small $\Lambda$
 it will make sense to restrict the gauge group to gauge
 transformations which leave a particular horizon volume invariant.
 The other gauge transformations will be viewed as physical
 operations, which act on the Hilbert space of the
 system\footnote{Indeed, as we have defined the Hilbert space
 $H_{\infty}$, in the previous section, all gauge transformations
 under consideration act on it. The distinction is between unitary
 maps which leave a given causal net of spaces embedded in
 $H_{\infty}$ invariant, and those which do not.} . Thus, we claim
 that for very large $N$, one can already begin to distinguish the
 special observables of the $N\rightarrow\infty$ boundary theory.
 In a sense, the large classical subsystems in a given horizon
 volume are acting like a Higgs field which picks out a particular
 gauge frame.
 
 There will clearly be corrections to such an approximate picture.
 However, not all corrections will disturb the
 approximate treatment of large subsystems by conventional quantum
 mechanics. We conjecture that as long as one does measurements
 that involve only a small fraction of the entropy of the universe,
 one does not have to worry about the conceptual issues of working
 with a finite system and doing measurements over finite time
 intervals. Only if one wanted to enquire into the nature of the
 exact quantum state of the Hawking DeSitter gas would one find
 oneself performing operations that were physically ambiguous.
 Since the Hawking temperature is very low, the uncertainty
 introduced into ordinary measurements by our lack of knowledge of
 the state of the horizon is exponentially suppressed as
 $N\rightarrow\infty$.

 \section{\bf Relation to a proposal of Witten}

 In his talk at Strings 2001 \cite{edtalk}, E. Witten proposed
 observables for spacetimes that are AsDS in the past, or the
 future, or both. His proposal was described in the standard
 global coordinate system for DeSitter space \eqn{DSmet}{ds^2 =
 -dt^2 + R^2 \cosh^2 (t/R)d\Omega^2,} in which the manifold is
 realized as a $d-1$ sphere that contracts from infinite radius to
 a finite radius $R$ and reexpands to infinity, as $t$ ranges over
 $[-\infty,\infty ]$. The boundary of this manifold is the union
 of the infinite spheres at past and future infinity. Witten
 proposes to define asymptotic states on these two boundaries (or
 on the union of the DS boundary at future infinity, and a Big Bang
 singularity in the past, for the case of an AsDS Big Bang
 cosmology). The attraction of this proposal is that it appears to
 define an S-matrix that is invariant under all diffeomorphisms of
 an AsDS spacetime. Witten claims that semiclassical analysis of
 this proposal leads to the conclusion that there are an infinite
 number of asymptotic states. He argues that this could be
 compatible with the claim of a finite number of states for AsDS
 spaces if the S-matrix he defines actually has finite rank in the
 infinite dimensional space of asymptotic states.

 Let us first assume that Witten's analysis is correct and inquire
 into the interpretation of this finite rank matrix. It appears to
 us that the procedure of modding out by the zero modes does not
 lead to a unique S-matrix. Unlike the case of BRST quantized
 gauge theories, the metric on the space of asymptotic states is
 positive definite (Witten claims to be analyzing diffeomorphism
 invariant physical states of quantum gravity perturbatively
 quantized around a DS background). Thus, it is impossible for
 different versions of the finite dimensional S-matrix to be
 equivalent to each other. There is no procedure for extracting it
 which is invariant under unitary transformations in the space of
 asymptotic states. That is, even if, using some conventions, we
 define a basis in which Witten's S-matrix is block diagonal, with
 one block being a finite dimensional unitary matrix $S_W^N$ and
 the others vanishing, we can, by performing a unitary
 transformation in the big space, turn this into $V S_W^N W$, so
 that the finite dimensional S-matrix is ambiguous. If this is
 the case, then we suspect that the ambiguity will turn out to be
 equivalent to the one arising in our prescription from the absence
 of an invariant way to choose a particular cosmological horizon.
 We have already explained why we think that this ambiguity
 actually reflects the correct physics of DS backgrounds via the DS
 Complementarity Principle.

 However, we also believe that the evidence for Witten's claim that
 the theory has an infinite number of states in the semiclassical
 approximation is less than compelling. One way to
 semiclassically quantize General Relativity in DS backgrounds is
 by analytic continuation of semiclassical Euclidean path integrals
 on the sphere. This analysis was presented in \cite{tbfolly}.
 General results of quantum field theory in curved spacetime relate
 these path integrals to gauge fixed free massless spin two fields,
 plus matter fields in the static patch of some particular
 cosmological horizon \cite{chiarabd}. Furthermore, the gauge fixing
 procedure instructs us to integrate over the DS group, which is
 part of the group of diffeomorphisms. Since the DS group maps
 every horizon patch into every other one, we are instructed to
 treat all information outside the horizon as a gauge copy of
 information inside the horizon. This is a strong argument in favor
 of the DSCP.
 The result of Euclidean functional integrals is the thermal state
 for the timelike Killing vector of the static patch. 
 We can construct other (impure) states by analytically continuing
 correlation functions of the form 
 $<O^{\dagger} (+) \phi (z_1) ... \phi (z_n) O(-)>$, where we have
 used the gauge fixing procedure to place a BRST invariant operator
 and its conjugate at the north and south poles of the coordinate
 we will analytically continue to Lorentzian time.  These give
 density matrices of the form $O^{\dagger} \rho O$, where $\rho$ is the
 thermal state. By
 construction, each of these gives rise to a DS invariant state in
 the sense of Witten. It is not clear to us whether all of the
 states described by Witten can be constructed in this way, but it
 would seem odd to find the correspondence between Euclidean and
 Lorentzian signature quantum field theory breaking down at this
 level.
 Assuming the two constructions are equivalent, we can inquire into
 the origin of the apparent infinite number of states in the
 construction. Since the static patch of DS space has finite
 volume, it is clear that this is a UV infinity. Thus, from the
 point of view of the static patch, the infinite number of states
 comes from the region where we do not trust the semiclassical
 approximation.
 
 How is this compatible with the analysis in the global coordinates
 \Ref{DSmet}? There the infinity can be viewed as coming from
 infinite numbers of well separated wave
 packets on the infinite radius spheres in the future and/or past.
 Since the spheres have infinite radius, we do not have to go to
 asymptotically high energy to localize excitations. So
 here we seem to establish the existence of an infinite number of
 states without invoking UV degrees of freedom for which the
 semiclassical approximation breaks down. We believe the key to
 understanding the consistency of the two analyses comes again from
 the singularity theorems of General Relativity. That is, we
 believe that initial conditions in global coordinates which appear
 to violate the Bekenstein-Hawking bound for DS space will lead to
 solutions with singularities. The physical mechanism for this is
 that the background DS evolution squeezes all matter into a finite
 radius $R$ at global time $t=0$. Thus, even very dilute matter on
 the sphere at infinity will evolve into (must have come from in
 the case of future infinity) matter whose gravitational field
 cannot be neglected.

 There appear to be two possible interpretations of these
 singularities. Some form of Cosmic censorship might be valid in
 AsDS spacetimes, in which case these initial conditions could be
 classified in terms of black holes in DS space. They would then
 have only finite entropy. On the other hand, some or all of the
 initial conditions violating the bound might have to be thrown
 away because they produced singularities that were unacceptable ({\it
 cf. } \cite{gubser}).Again, the question of infinite numbers of physical
 states would seem to depend on the behavior of the theory in regimes
 where the semiclassical approximation breaks down.
 In our opinion, the most likely conclusion is that the system
 has only a finite number of physical states and a truly
 gauge invariant S-matrix does not exist. This reflects the
 semiclassically verifiable conclusion that observers in different
 horizon patches will see different semiclassical physics, combined
 with the DS Complementarity Principle, which asserts that all of
 these observations are measuring operators defined in a single
 finite dimensional Hilbert space. Of course, it is only in the
 limit of very large dimension (very small cosmological constant)
 that we expect any of the physics to have a semiclassical
 description.

\section{Discussion - the relation to string theory}

What is the relation of all of this to string theory or M-theory 
(which we use in the sense of the theory underlying the various
semiclassical expansions embodied in perturbative string theory and 11D SUGRA)?
We believe that M-theory has various incarnations which, loosely
speaking, depend on asymptotic boundary conditions in spacetime.
There is no background independence in the sense that most 
string theorists have assumed in recent years.  That is
, not all of the versions of M-theory can be thought of as
different representations of the same operator algebra in Hilbert
space\footnote{This is the most general way to phrase what we mean
by different vacuum states of the same theory in quantum field theory.}
.   M-theory in asymptotically AdS spaces is described quantum
mechanically by conformal quantum field theory.  M-theory in asymptotically
flat spaces is described by some as yet undiscovered quantum operator algebra
which naturally reproduces the high energy black hole spectrum
\footnote{In certain cases it can in principle be constructed as the large $N$
limit of Matrix Theory.}.   M-theory in linear dilaton backgrounds is Little
String Theory.   M-theory in $\Lambda = 0 $ FRW spacetimes  may have a 
a description in terms of the same operator algebra as that of asymptotically
flat spaces, but also requires a dual description in terms of simple
operators at the Big Bang.  The gauge invariant information in this theory
is encoded in the S-matrix between these two descriptions.
M-theory in AsDS spaces lives in a finite dimensional Hilbert space, {\it etc.}.

Our most far reaching conjecture would be that somehow all of these
different versions of M-theory could be realized in terms of a set 
of axioms for nets of Hilbert spaces of the type we have 
discussed here.  The different classes of M-theory would correspond to
different asymptotic conditions on the net.   Our ability to encode
spacetime geometry in terms of Hilbert spaces associated with a Planck 
lattice makes us optimistic that such a construction will be possible.
What is missing is a formulation of the dynamical laws directly
in terms of the net of Hilbert spaces.

This philosophy leads us to be skeptical of attempts to find
{\it string theory models of DS space}.  If we are
correct, both asymptotically flat string theory, and DS space,
will be realized as special cases of the same underlying laws,
but string theory will have too many degrees of freedom
to describe DS space.  An interesting possibility\footnote{Suggested to
us by E.Silverstein.} is the construction of a metastable DS
vacuum in models that are {\it e.g.} weakly coupled string
theory in something like asymptotically flat space.  Then the conventional
stringy S-matrix might contain information about the
DS ``resonance''.   If Newton's constant is asymptotically
constant, this attempt is likely to fail and the putative DS vacuum will remain
forever shrouded behind a black hole \cite{gf}.   However,
in models where Newton's constant asymptotically rolls to zero,
some progress might be made.
   
 \section{Appendix - quantum mechanics and cosmology}
 
 The idea of using quantum mechanics to describe the entire
 universe has been known to inspire unease in the breasts of some
 of our most eminent physicists. The basic problem is that the
 physical interpretation of quantum mechanics appears to depend
 crucially on the ideas of probability theory, and the operational
 definition of probability requires us to imagine the possibility
 of doing a measurement an infinite number of times under exactly
 equivalent conditions. On the other hand, the evolution of the
 universe occurs only once. Furthermore, the concept of
 measurement requires us to separate the universe into system and
 apparatus, and this may not be possible, even in principle, under
 all cosmological conditions.
 
 We believe that there is a formulation of the principles of
 quantum mechanics which ameliorates this philosophical distress,
 without perhaps removing it entirely. It is essentially the
 Quantum Logic interpretation of Quantum Theory proposed by Von
 Neumann (and immediately dismissed by Bohr as a mathematical
 irrelevancy). Here we will present a brief review of this
 interpretation with a few linguistic twists.
 The essential observation is that classical logic can be
 reformulated in terms of a $C^*$-algebra \footnote{For simplicity
 of exposition, we will imagine a physical system with only a
 finite number of states, so we really mean just a finite
 dimensional matrix algebra.}. Any question about a physical
 system can be turned into a yes no question. Thus, the statement
 that a certain variable has the value $5.2$ is equivalent to the
 questions of whether or not it takes on any one of its allowed
 values, and whether $5.2$ is among them. It is well known that
 the logical relations between any finite number of yes/no
 questions are equivalent to the (Boolean) algebraic relations
 between a maximal set of commuting projections in a finite
 dimensional Hilbert space. It is further obvious that we want to
 define a general observable in such a system as a real linear
 combination of the projection operators. We introduce complex
 linear combinations for the purely mathematical convenience of
 being able to solve general algebraic equations involving
 observables (it would be more satisfying to have a physical
 motivation for introducing complex numbers). 

 Thus, thinking purely
 classically, a physical system is related to the Hermitian
 elements of a $C^*$-algebra of operators in a Hilbert space.
 The quantum logician then makes the observation that this
 classical logic has chosen a special basis in the Hilbert space,
 or equivalently a special maximal abelian subalgebra. Being
 mathematically minded, she asks what the choice of a state of the
 system (a choice of the answers to all of the classical logician's
 yes/no questions) implies for all of the other Hermitian operators
 which do not commute with the special abelian subalgebra. She
 quickly realizes that each such state defines a probability
 distribution for any other maximal abelian subalgebra, and further
 that any such probability distribution for a given maximal abelian
 subalgebra is equivalent to a choice of classical state for {\it
 some other} maximal abelian subalgebra. Our quantum logician is
 one of those fortunate people who has both mathematical and
 physical intuition and invents the notion of {\it complementary
 observations}. That is she declares that all of the Hermitian
 operators in the Hilbert space represent possible observations on
 the physical system but that some measurements interfere with the
 results of others.

 In particular, one must contemplate the change of the state of the
 physical system with time, an operation which, {\it even for the
 classical logician} is described by a unitary operator which does
 not commute with with his favored abelian subalgebra. The quantum
 logician identifies Hermitian functions of this unitary operator as
 candidate observables which will be complementary to the classical
 logicians preferred measurements.

 If one can imagine an alternative history in which Boolean logic
 and the theory of operator algebras was developed previous to the
 invention of classical mechanics, one can imagine the seventeenth
 century quantum logician pondering the question of whether some of
 Mr. Newton's observable quantities might be complementary to each
 other. Presuming her to be very long lived, and noting that, as
 one of the few women in theoretical physics at the time, she would
 have been particularly sensitive to E. Noether's famous result
 connecting symmetries (particular operations on or changes of
 state) of classical mechanical systems with conservation laws
 (particular observables), one can imagine our quantum logician
 jumping to the absurd conclusion that position and momentum were
 complementary variables. Long after her speculative ideas were
 rejected by the male dominated physics community
 she would be awarded a posthumous Nobel Prize when
 Jordan realized that her mathematical speculations provided the
 key to the mysterious behavior of atomic systems.

 The point of this fairy story is that the idea of complementarity
 is in fact quite natural from the point of view of classical
 logic, once it is embedded in an appropriate mathematical
 framework. There is nothing in quantum mechanics which is
 intrinsically probabilistic until one insists on measuring
 complementary variables. Much of our unease with quantum
 mechanics stems from the fact that, as a consequence of our own
 physical characteristics (the fact that the typical classical
 action in processes familiar to us before the advent of technology
 is much larger than Planck's constant), we mistook certain
 complementary variables for elements of the same abelian
 subalgebra.

 Given a solution $|\psi (t)\rangle$ of the Schrodinger equation for any
 quantum system, we can construct a complete commuting set of (time
 dependent) observables, which remain sharp throughout the
 evolution. These are simply the projector on the time dependent
 state of the system, and any complete commuting set of orthogonal
 projectors. Normally, we would reject this statement as a
 mathematical irrelevancy. Here are some of the reasons why:
 \begin{itemize}
 \item For a system with a time independent Hamiltonian, it is often
 convenient to introduce fundamental variables which allow us to
 simply describe the high energy (and thus short time) behavior of
 the system. In particular, the familiar $p$'s and $q$'s of
 classical mechanics are appropriate for systems whose high energy
 behavior is described by a Gaussian fixed point. The sharp
 observables above are not simply related to the Gaussian
 variables.
 \item For systems which are under true experimental control, we
 like to do repeated experiments with different initial conditions.
 The sharp observables defined above depend on the initial
 condition and thus do not provide a convenient way of
 characterizing all of the experiments we do on the system in a
 universal framework.
 \item A related problem is that the the measurements we actually
 make with external probes on an isolated system have no simple
 relation to the sharp observables.
 \end{itemize}
 The acute reader will have realized that none of these objections
 apply to the discussion of the universe as a quantum system. The
 conventional objection that the universe only happens once is
 precisely the reason that the second objection to sharp
 observables is irrelevant. Observation shows us that the
 universe does not have a time independent Hamiltonian, and so the
 first objection is irrelevant as well. Indeed, our discussion of
 the Big Bang in the body of the paper suggests that at very early
 times the state of the system may be characterized in terms of an
 integrable CFT. At very late times, the high energy behavior is
 surely dominated by black hole states, or if we live in an AsDS
 space, by states near the horizon. Neither of these regimes seems
 to have a standard classical description.
 The final objection to the sharp observables for the universe has
 more to do with our own limitations than with limitations of the
 applicability of quantum theory to the description of the
 universe. The conditions for the existence of independent (let
 alone intelligent) complex systems which can do measurements on
 pieces of the universe without affecting other parts of it, are
 very special. In the case of an asymptotically FRW universe such
 systems can exist in the asymptotic future but are unlikely to
 exist near the Big Bang. If the universe is AsDS such systems do
 not even exist in the asymptotic future (eventually everything
 either collapses into a black hole or is thermalized by the
 background DeSitter radiation). It is not surprising then that
 the natural quantities measured by these approximately isolated
 systems are not compatible with the classical state of the
 universe as defined by $| \psi (t)\rangle$. Mathematically, we
 can view the universe as evolving deterministically in a
 classical state determined as above in terms of projectors on and
 orthogonal to its wave function. But these classical observables
 are not measurable by the kinds of apparati that can be
 constructed out of subsystems to which the approximate notion of
 locality applies. The probabilistic nature of the universe as we
 view it is a characteristic of the nature of the things we can
 measure rather than of the universe itself.

 %\noindent {\bf Acknowledgements:}
 \acknowledgments
 We would like to thank, D.Gross, L.Motl, L.Susskind, and particularly R.Bousso for
 illuminating conversations about cosmology and the holographic
 principle. The work of T.Banks was supported in part by the Department
 of Energy. The work of W.Fischler was supported in part by the Robert
 A. Welch Foundation and NSF Grant PHY-0071512.
 %
 %%%%%%%%%%%%%%%%%%%%%%%%%%%%%%%%%%%%%%%%%%%%%%%%%%%%%%%%%%%%%%%%%%%%%%%%%%%%
 % REFERENCES %
 %%%%%%%%%%%%%%%%%%%%%%%%%%%%%%%%%%%%%%%%%%%%%%%%%%%%%%%%%%%%%%%%%%%%%%%%%%%%
 
 \end{document}